# Modeling of Microbubble Enhanced High Intensity Focused Ultrasound

Aswin Gnanaskandan[*], Chao-Tsung Hsiao[‡], Georges Chahine[†],

---

[*] Corresponding author, Research Scientist, DYNAFLOW, INC., 10621-J Iron Bridge Road Jessup, Maryland 20794, Tel: (301)604-3688, Email: aswin@dynaflow-inc.com
[‡] Principal Research Scientist, DYNAFLOW INC., 10621-J Iron Bridge Road Jessup, Maryland 20794, Tel: (301)604-3688, Email: ctsung@dynaflow-inc.com
[†] President, DYNAFLOW, INC., 10621-J Iron Bridge Road Jessup, Maryland 20794, Tel: (301)604-3688, Email: glchahine@dynaflow-inc.com

**Abstract**

Heat enhancement at the target in a High Intensity Focused Ultrasound (HIFU) field is investigated by considering the effects of the injection of microbubbles in the vicinity of the tumor to be ablated. The interaction between the bubble cloud and the HIFU field is investigated using a three-dimensional numerical model. The propagation of non-linear ultrasonic waves in the tissue or in a phantom medium is modeled using the compressible Navier-Stokes equations on a fixed grid, while the microbubbles dynamics and motion are modeled as discrete singularities, which are tracked in a Lagrangian framework. These two models are coupled to each other such that both the acoustic field and the bubbles influence each other. The temperature rise in the field is calculated by solving a heat transfer equation applied over a much longer time scale. The compressible continuum part of the model is validated by conducting HIFU simulation without microbubbles and comparing the pressure and temperature fields against available experiments. The coupled Eulerian-Lagrangian approach is then validated against existing experiments with a phantom tissue. The various mechanisms through which microbubbles enhance heat deposition are then examined in detail. The effects of the initial void fraction in the cloud are then sought by considering the changes in the attenuation of the primary ultrasonic wave and the modifications of the enhanced heat deposition in the focal region. Finally, the effects of the microbubble cloud size and its localization in the focal region are shown and the effects of these parameters in altering the temperature rise and the location of the temperature peak are discussed.

# Introduction

High Intensity Focused Ultrasound (HIFU) uses the focused energy of sound waves to elevate temperature locally, causing thermal ablation of tissues. HIFU therapy has kindled a great interest in the scientific community because of its non-invasive nature and its potential to treat deep-seated cancers such as those in the liver and brain (Kennedy 2005, Kennedy et al. 2004). A major impediment with the current use of the HIFU technique to efficiently treat deep-seated cancer is the long treatment time whilst using high intensity insonation for deeper penetration, as higher intensity sound waves may cause unwanted tissue damage along the waves' passage before reaching the targeted region. In order to reduce undesirable damage of surrounding tissues, the resting time between insonations to cool the pre-focal region has to be increased. It is therefore desirable to use means to generate higher temperature elevations locally in the target region while still using moderate intensity levels (100 – 1000 W/cm$^2$), which do not harm the tissue along the wave passage (Hariharan et al. 2007). Introducing microbubbles in the form of contrast agents has been shown to increase temperature levels in both *in vitro* (Kajiyama et al. 2010, Razansky et al. 2006) and *in vivo* experiments (Chung et al. 2012, Kaneko et al. 2005). However, the exact mechanism of how the microbubbles contribute to the heat increase is not known. Only some educated speculations were made due to the difficulty in both experimental measurements and computational modeling. Two mechanisms were proposed for heat enhancement viz. heating through acoustic emission and through viscous dissipation (Holt & Roy 2001).

Acoustic emission from the bubble oscillations arises because the bubble radiates acoustic energy. This becomes important primarily when the bubble undergoes large amplitude oscillations (i.e. inertial oscillations). Viscous damping arises primarily from viscous dissipation in the relatively thin viscous layer of host medium surrounding the bubble during its oscillations. The dissipation is directly proportional to the bubble interfacial velocity and the viscosity of the host medium. The relative contribution of these terms (viscous vs. acoustic) depends on a variety of parameters, which can be elucidated through numerical simulations. The main objective of the work presented here was to develop a numerical model to simulate accurately HIFU both in the absence and presence of microbubbles, verify its validity, and apply it to understand the effect of bubble parameters on heat enhancement.

One of the major difficulties in characterizing experimentally a HIFU field is due to the high intensity levels applied to the irradiated region (Hynynen & Clement 2009). Measurements are restricted by potential damage to the sensors and by the small lateral dimensions of the focal area. Detailed measurements are thus performed at low driving amplitudes and the results are extrapolated to higher intensities, thus neglecting important nonlinear effects. Numerical modeling is therefore required to address the high intensity conditions (Canney et al. 2008). A commonly used model for nonlinear acoustics in HIFU applications is based on the Westervelt equations (Hamilton & Blackstock 1998) or the KZK equations (Bakhvalov et al. 1987). These nonlinear acoustics equations provide the acoustic pressure field and are coupled with a bio-heat transport equation to predict ultrasound heat deposition (Gheshlaghia et al. 2015, Pennes 1948). However, coupling between the acoustic field obtained with KZK equations and bubble dynamics has been limited to one-way interaction, where the modification of the ultrasound field by the bubbles is ignored, thus compromising the accuracy of high intensity ultrasound simulations.

Conventional numerical approaches for modeling bubbly two-phase mixture use continuum flow models with ensemble-averaging techniques (Zhang & Prosperetti 1994, Biesheuvel & Wijngaarden 1984). These models solve the mixture flow using a fixed grid accounting for averaged bubble dynamics effects and ignoring scattering and local nonlinear effects due to the bubbles. In such approaches, interactions between neighboring bubbles are not directly considered and this interaction is indirectly accounted for through the averaged two-phase flow field (Grandjean et al. 2012, Ando et al. 2011, Kameda & Matsumoto 1996). In this study, we consider a more direct two-way coupling of the acoustic field and the bubble dynamics using an Eulerian-Lagrangian approach (Maeda et al. 2017, Ma et al. 2015a, Chahine et al. 2014, Okita et al. 2013). The bubbles are tracked in a Lagrangian fashion, while the acoustic and thermal fields are resolved using an Eulerian fixed grid. Non-uniformities in the flow field are taken into account through a surface averaging method in which the flow quantities driving the bubble dynamics (pressure, velocity, and their gradients) are obtained through an arithmetic average of these mixture properties at six polar points on the bubble surface. The coupling is two-way, i.e. the acoustic field drives the bubble dynamics and the bubble behavior affects the acoustic field dynamically. Heat deposition resulting from the high intensity acoustic waves' effects on the viscous tissue and from bubble oscillations, is modeled using a bio-heat transport equation (Hariharan et al. 2007, Huang 2002). Most continuum models assume that the two phase mixture is homogenous (Gnanaskandan & Mahesh 2015, Kinzel et al. 2009, Singhal et al. 2002), ignoring slip velocity between the phases. However, experimental observations and numerical models accounting for slip velocity have shown a potential for large bubble-liquid relative motion such as strong micro-streaming under ultrasonic horns (Chahine et al. 2016, Mannaris & Averkiou 2012). Therefore, accurate description of the microbubbles' motion and interaction is very important for microbubble enhanced HIFU applications since this affects the amount of heat deposition. Our Eulerian-Lagrangian method was successfully applied to several problems including complex geometries (Hsiao et al. 2017, Ma et al. 2015b, Chahine et al. 2014, Hsiao & Chahine 2012) using a pseudo-compressible continuum model solving the Navier Stokes equations. In the present study, we further develop the Eulerian-Lagrangian approach considering full compressibility of both gas and liquid, capturing shock waves, and including heat equations to derive the temperature field.

The paper is organized as follows. We first present the governing equations and the numerical methodology where the details of the Eulerian and Lagrangian approaches are explained. We then present the model used to solve the heat transfer equation and to approximate the heat source terms used in this equation. We then validate the methodology in the absence of bubbles in both water and experimentally used phantom tissue. Finally, a HIFU experiment with microbubbles, available in the literature (Kajiyama et al. 2010), is simulated and the heat enhancement obtained due to the bubbles is compared to the experiments. The contribution of various heat source terms is then discussed. This is followed by the presentation of a parametric study on the effects of void fraction and the localization of the bubbles. The paper is then concluded with the presentation of a brief summary and of the main conclusions.

## Governing equations and Numerical Methodology

### Compressible Flow Solver

The compressible flow solver, 3DYNAFS-COMP, describing acoustic wave propagation through a two-phase medium and the associated acoustic streaming, solves the following governing equations for conservation of mass, momentum, and energy in a fixed reference frame:

$$\frac{\partial \rho_m}{\partial t} + \nabla \bullet (\rho_m \mathbf{u}) = 0,$$
$$\frac{\partial \rho_m \mathbf{u}}{\partial t} + \nabla \bullet (\rho_m \mathbf{uu} + p_m \mathbf{I} + \boldsymbol{\sigma}) = 0, \quad (1)$$
$$\frac{\partial \rho_m E_m}{\partial t} + \nabla \bullet ((\rho_m E_m + p_m)\mathbf{u}) + \boldsymbol{\sigma}.\mathbf{u} + q) = 0.$$

Here $\rho_m$, $E_m$, and $\mathbf{u}$ are the mixture density, total energy, and velocity respectively. The mixture density is defined using the mixture components' densities and volume fractions (the components being the tissue or phantom medium and the gas inside the bubbles) and is given by:

$$\rho_m = \sum_i \rho_i \alpha_i, \quad \text{where} \quad \sum_i \alpha_i = 1. \quad (2)$$

The total energy is also given by:

$$E_m = e_m + 0.5\mathbf{u}^2, \quad \text{where} \quad \rho_m e_m = \sum_i \rho_i e_i \alpha_i. \quad (3)$$

The shear stress tensor σ and the heat conduction term $q$ are computed using a mixture viscosity defined as:

$$\mu_m = \sum_i \mu_i \alpha_i. \tag{4}$$

In Eq. (1) $p_m$ denotes the mixture pressure and is obtained from a mixture equation of state (EOS). Such an EOS is derived by considering that the background medium obeys a stiffened equation of state given by:

$$p = (\gamma - 1)\rho e - \gamma \pi, \tag{5}$$

and that the gas inside the bubbles obey an ideal gas law also expressed in a stiffened equation form with $\pi = 0$.

The mixture EOS (Pelanti & Shyue 2014) is then given by

$$\begin{aligned} p_m &= (\Gamma - 1)\rho_m e_m - \Gamma \Pi, \\ \Gamma &= 1 + \frac{1}{\sum_i \dfrac{\alpha_i}{\gamma_i - 1}}, \\ \Pi &= \frac{\Gamma - 1}{\Gamma} \sum_i \alpha_i \frac{\gamma_i \pi_i}{\gamma_i - 1}. \end{aligned} \tag{6}$$

When one of the component volume fractions reduces to zero, this form of the governing equations (1) through (6) automatically reduces to a single component Navier-Stokes equation closed with a stiffened EOS. These equations are solved using a fully conservative higher order Monotonic Upwind Scheme for Conservation Laws (MUSCL) scheme (Van Leer & Woodward 1979) and an approximate Riemann solver (Kapahi et al. 2015, Colella 1985). Since we consider only two components viz. tissue phantom/water and gas bubbles, $\alpha_1 + \alpha_2 = 1$, and hence the only unknown required to close the system of equations is the volume fraction of gas bubbles, denoted hereafter as $\alpha$, which is obtained from the knowledge of the local spatial distribution of the microbubbles. We compute this by tracking the bubbles using the Discrete Singularity Method (DSM) (Choi et al. 2007, Hsiao & Chahine 2012).

**Bubble Modeling through the Discrete Singularity Method**

In a known pressure and velocity field, a microbubble assumed to remain spherical and of a radius $R(t)$, can be tracked in a Lagrangian fashion using a bubble dynamics equation such as the Keller-Herring equation (Prosperetti & Lezzi 1986), and a bubble motion (translation) equation (Johnson & Hsieh 1966). The bubble radius time evolution follows the differential equation:

$$(1-\frac{\dot{R}}{c_m})R\ddot{R}+\frac{3}{2}(1-\frac{\dot{R}}{3c_m})\dot{R}^2 =$$
$$\frac{1}{\rho_m}(1+\frac{\dot{R}}{c_m}+\frac{R}{c_m}\frac{d}{dt})\left[p_v+p_g-p_m-\frac{2\gamma}{R}-4\mu_m\frac{\dot{R}}{R}-\frac{4}{3}G\left[1-\left(\frac{R_0}{R}\right)^3\right]\right] \quad (7)$$
$$+\frac{(\mathbf{u}-\mathbf{u}_b)^2}{4},$$

where $\rho_m$, $c_m$, and $\mu_m$ are the density, sound speed, and viscosity of the surrounding medium. $\dot{R}$ and $\ddot{R}$ represent the bubble interface velocity and acceleration respectively. $p_v$ and $p_g$ are the vapor and gas pressure inside the bubble and $p$ is the surrounding surface averaged pressure driving the bubble dynamics. $\gamma$ is the surface tension, and $G$ is the shear elasticity of the medium surrounding the bubble.

The bubble motion equation can be written as follows accounting for viscous drag, lift, and slip velocity between the bubble (of velocity $\mathbf{u_b}$) and the mixture (of velocity $\mathbf{u}$),

$$\begin{aligned}
\frac{d\mathbf{u}_b}{dt} = &-\frac{3}{\rho_m}\nabla p + \frac{3}{4}\frac{C_D}{R}(\mathbf{u}-\mathbf{u}_b)|\mathbf{u}-\mathbf{u}_b| + \\
&\frac{3\dot{R}}{R}(\mathbf{u}-\mathbf{u}_b) + \frac{3}{2\pi}\frac{C_L}{R}\sqrt{\frac{\mu_m}{\rho_m}}\frac{(\mathbf{u}-\mathbf{u}_b)\times\boldsymbol{\omega}}{\sqrt{|\boldsymbol{\omega}|}},
\end{aligned} \quad (8)$$

where $C_D$ is the drag coefficient (Haberman & Morton 1953), $C_L$ is the bubble lift coefficient (Saffman 1965), and $\boldsymbol{\omega}$ is the local vorticity. The bubble motion and its volume evolution are obtained by integrating Eqns (7) and (8) over time with an explicit fourth order Runge-Kutta scheme. The surface average quantities are obtained through an arithmetic averaging of the mixture pressures and velocities at six polar points on the bubble surface.

**Void Fraction Computation**

Once all bubbles' instantaneous sizes and locations are computed, it is necessary to communicate this information back to the compressible flow solver as the local void fraction. This is achieved by computing an effective void fraction derived from the contribution of each bubble to its surrounding computational cells using a Gaussian distribution as illustrated in Figure 1. The Gaussian distribution scheme computes $v_{i,j}$, the bubble $j$'s volume contribution to the void fraction computation in cell $i$, using:

$$v_{i,j} = V_j^b e^{-\frac{|\mathbf{x}_i - \mathbf{x}_{0,j}|^2}{\lambda R_j^2}}. \quad (9)$$

Here $V_j^b$ is the volume of bubble $j$, $\mathbf{x}_i$ is the coordinate center of cell $i$ and $\mathbf{x}_{0,j}$ is that of the center of bubble $j$, and $\lambda$ is the characteristic radius of influence of the bubble. However, this does not guaranty that the total volume of the bubbles is conserved.

Therefore, a cell-volume-weighted normalization scheme is adopted to normalize the volume contribution, i.e.

$$\bar{v}_{i,j} = \frac{v_{i,j} V_i^{cell}}{\sum_{k}^{N_{cells}} v_{k,j} V_k^{cell}} V_j^b, \tag{10}$$

where $V_i^{cell}$ is cell $i's$ volume. Since each bubble only contributes its volume to a limited number of nearby cells due to the Gaussian decay, the normalization is computed only for those cells of total number Ncell, which are influenced by the bubble $j$. To compute the void fraction for the cell $i$ we then sum up the contributions of all bubbles within the "influence range" and divide it by the cell volume, i.e.

$$\alpha_i = \sum_{j=1}^{N_i} \frac{\bar{v}_{i,j}}{V_i^{cell}} = \sum_{j=1}^{N_i} \frac{v_{i,j} V_j^b}{\sum_{k}^{N_{cells}} v_{k,j} V_k^{cell}}, \tag{11}$$

where $N_i$ is the number of bubbles which are in the "influence range".

**Model for Focused Ultrasound**

The ultrasound source emission is modeled as a pressure distribution with phasing imposed on the inlet boundary (z=0) (Canney et al. 2008),

$$p(z=0, r, t) = p_0 \sin\left[2\pi f_0 \left(t + \frac{r^2}{2cF}\right)\right], \tag{12}$$

where $p_0$ and $f_0$ are the amplitude and frequency of the ultrasound and $r$ and $F$ are the radius and focal length of the transducer. Such a boundary condition produces a spherically focused wave with a focal length F and radius r.

### Modeling of Heat Deposition

The insonation time during clinical operation is of the order of one second. This is $10^6$ times the period of the acoustic waves. Since the time accurate acoustic field CFD computations are limited to tens of cycles only, we use a decoupled approach (Okita et al. 2013, Hariharan et al. 2007, Huang et al. 2004), where we develop the flow field solution using the Eulerian-Lagrangian approach described above, then we separately solve a heat transfer equation, which addresses the longer time

$$\rho C_p \frac{\partial T}{\partial \tau} = K \nabla^2 T + q_{US,AC} + q_{VIS}. \tag{13}$$

Here $\rho$, $C_p$, and $K$ are the density, specific heat, and thermal conductivity of the medium, $T$ is the temperature and $\tau$ is time. The heat sources, $q_{US,AC}$ and $q_{VIS}$, in this equation are due to heating from the primary ultrasound source and acoustic emission from bubble oscillations, and heating due to viscous damping of the bubbles respectively. The source terms, $q_{US,AC}$ and $q_{VIS}$, are obtained as time-averaged values computed during the solution of the Eulerian-Lagrangian two-phase problem. These time-averaged sources are then used to drive the heat equation. The time averages are computed over at least 10 cycles, after the initial transients are removed and after the primary ultrasound wave reaches the geometric focus.

The heat source $q_{US,AC}$ due to the primary ultrasound absorption and acoustic emissions of the bubble is given by

$$q_{US,AC} = \mu_b \varepsilon_{kk}^2 + 2\mu \left( \varepsilon_{ij} - \frac{1}{3} \varepsilon_{kk} \delta_{ij} \right)^2, \tag{14}$$

where $\mu$ and $\mu_b$ are the shear and bulk viscosities of the mixture medium and $\varepsilon$ is the strain rate tensor. The viscosity of the medium is often not known directly from measurements, but can be estimated from the absorption coefficient, $\Omega$, (Canney et al. 2008), using Stokes' law of sound attenuation,

$$\mu = \frac{3}{2}\left[\frac{\Omega \rho c^3}{\omega^2} - \mu_b\right], \tag{15}$$

where $\omega$ is the angular frequency. We further assume that the bulk viscosity of the medium, $\mu_b$, is thrice the dynamic viscosity, $\mu$ (Holmes et al.).

The heat addition due to the viscous damping of a single bubble is given by (Okita et al. 2013, Holt & Roy 2001):

$$q_{vis}^b = \left(4\pi R^2\right) \cdot \left(4\mu \frac{\dot{R}^2}{R}\right). \tag{16}$$

In order to compute the viscous heating contribution of each bubble to a given control volume, we follow a procedure similar to that of void fraction computation described in Eqns. (9-11). To compute the viscous heat contribution for the cell $i$ we sum up the contributions of all bubbles within the "influence range" as described in Section 'Void Fraction Computation' and divide it by the cell volume :

$$q_{VIS} = \sum_{j=1}^{N_i} \frac{q_{vis}^b V_j^b}{\sum_{k=1}^{Ncells} v_{k,j} V_k^{cell}}. \tag{17}$$

## Results and Discussion

All the results presented in this paper are obtained through axisymmetric simulations. When bubbles are present, they are distributed in the three-dimensional cylindrical sector $[0, \theta_{sector}]$. The void fraction contribution of each bubble inside this three dimensional volume is obtained using (11). This three-dimensional void fraction distribution, $\alpha(r,z,\theta)$, is then used to obtain an averaged axisymmetric void fraction distribution using

$$\alpha_{axi}(r,z) = \frac{1}{2\theta_{\text{sector}}} \int_0^{\theta_{\text{sector}}} \alpha(r,\theta,z)d\theta. \tag{18}$$

A CFL number of 0.2 is used for all the simulations. The grids used in the simulations are primarily determined by the wavelength of the imposed ultrasound. It was ascertained through numerical experiments that at least 20 points per wavelength is needed to capture the wave propagation with negligible dissipation. This resolution also ensures that higher harmonics can also be captured partially, although there is no guaranty that numerical dissipation will not affect the higher harmonics.

A grid sensitivity study carried out for a focused ultrasound at a frequency of 1.1 MHz and an excitation amplitude of 0.01 MPa is shown in Figure 2. The radius of the transducer is considered to be 5mm and the focal length is 10mm. Three grids are used with consecutively finer resolution per wavelength. The first simulation is one where a wavelength is resolved using 20 grid points, the second where 40 grid points are used and finally 80 grid points are used. The pressure history obtained at the focus shows that the focal pressure values obtained using the three grids are very close to each other. There are practically no differences between the pressures for 40 points and 80 points. However, using only 20 points per wavelength results in an error in the amplitude and timing of the peak of about 7% (see Figure 2(b)). However, owing to the cost involved in resolving a wave using 80 grid points, we choose 20 grid points per wavelength to conduct the simulations and preliminary parametric studies considered in this paper.

**HIFU Simulation in Water without Microbubbles**

The fully compressible continuum solution method is first verified by computing the behavior and propagation of focused ultrasound waves in water in both linear and non-linear regimes. For the linear regime, the results are compared with the experiments of Huang (2002), where a 1.1 MHz transducer of radius 35 mm and focal length 60 mm is used to generate an acoustic wave with a pressure amplitude of $P_a$ = 0.01 MPa. We compare the focal scan data obtained along the axis with the numerical results in Figure 3(a). The focal scan showing the magnitude of the pressure peaks along the axis agrees well with the experiment indicating that the area in which the ultrasound is focused is predicted properly.

In Figure 3(b), we compare the pressure history obtained at the focal point for a 2.2 MHz transducer of 20 mm radius and 44 mm focal length at a pressure amplitude at the source of 0.29 MPa corresponding to the experiments of Canney et al. (2008). At this source pressure, non-linear steepening of the acoustic waves results in a shock wave at the focus, which is also captured well by the numerical simulations.

**HIFU Simulations in Agar Phantom without Microbubbles**

Next, we validate the numerical method for a HIFU procedure in a phantom tissue in the absence of bubbles. In the experiments of Huang et al. (2004), a phantom tissue made of Agar was subjected to focused ultrasound from a 1.1 MHz transducer of radius 35 mm and focal length 60mm. In the computation, the considered properties of the Agar are $\rho$ = 1,044 Kg/m$^3$, $\mu$ = 0.3 Pa s and shear elasticity is neglected. The resulting ultrasound field is depicted in Figure 4(a), where the acoustic waves are seen to focus at the geometric focal point, where the spherical waves converge The heat absorbed by the phantom tissue is obtained as a time-averaged quantity over 0.01 ms and is then used as source term in the heat equation. The time-averaged contours of heat release per unit time ($q_{US}$), depicted in Figure 4(b), shows the concentration of $q_{US}$ around the focal region. Note that the contours in Figure 4(b) are in logarithmic scale for clarity and that the highest value in the focus area is at least an order of magnitude higher than that in the pre and post focal regions.

Comparison of the temperature history between simulation and experiments at the focus and at one off-focus location is depicted in Figure 5. The ultrasound source is on for one second followed by a cooling time of 4s. The simulations results show reasonable agreement with the experiments indicating the accuracy of the physical model and the numerical method in simulating HIFU flow field. Increase in computational grid density should further improve the comparison.

### HIFU simulations in Polyacrylamide (PA) phantom with microbubbles

Finally, the model is applied to study HIFU in the presence of microbubbles. The simulations correspond to the *invitro* experiments with microbubbles of Kajiyama *et al.* (2010). The schematic of the experimental setup is shown in Figure 6. A transducer of radius 20 mm and focal length of 40 mm is used to insonate a phantom tissue made of Polyacrylamide gel at a frequency of 2.2 MHz. For the experiments with microbubbles, Levovist contrast agents solution of pre-determined void fraction is inserted inside a cylindrical space of radius 5 mm and depth 10 mm around the geometric focus inside the gel. It is not clear from the experimental setup if the geometric focus lies exactly at the geometric center of this cylindrical region. We will assume it to be so in the numerical simulations. The maximum diameter of the microbubbles in the experiment do not exceed 10 μm and the average diameter is 1.3 μm. For the numerical simulations, a uniform bubble size of 1.3 μm diameter is considered. The effect of polydispersity, not considered in this study, will be the subject of our follow up studies. The density of the gel is 1,060 Kg/m$^3$, the viscosity is 0.01 Pa s, and the shear elasticity is 0.1 MPa. In the experiments, the insonation time was 60 s and the peak intensity at the focus was 1,000 W/cm$^2$. Acoustic calculations are carried out for 0.1 ms with time-averaging for the source terms for the heat equation over 0.01 ms. The heat equation is then solved for 60 s with the derived source terms.

The temperature rise obtained from the simulations in the focal region is compared with the experimental data in Figure 7. The figure shows the temperature evolution for both no microbubble condition and in the presence of microbubbles with a void fraction, $\alpha = 10^{-5}$. Note that the exact location of the thermocouple measuring the temperature in the experiment is not known and that the reported error in the temperature measurement was large (±5 K). Hence in the numerical study, the temperature obtained along the axis at various locations (40 mm < Z < 45) mm are shown and compared with the experiment. The space averaged values in the same range are also shown. While the average temperature rise in the focal region without bubbles is approximately 3.5 K, in the presence of bubbles, a temperature rise of 10 K is obtained. Given that the uncertainty in the experimental temperature is about 5 K, the agreement between the simulations and experiment is reasonable, with the temperature ranges in each case captured and with significantly higher temperatures obtained in the presence of the bubbles. It is worthwhile to note that although only data along the axis is shown here, the maximum temperature can occur off-axis when bubbles are present. This is illustrated in Figure 8 as contours of temperature rise in the XZ plane. In the absence of bubbles (Figure 8(a)), the maximum temperature is obtained along the axis. However, when bubbles are present (Figure 8(b)), heat rise is dominant in the vicinity of the bubbles undergoing oscillations. This leads to high temperature regions away from the axis as well, since all the bubbles are not located on the axis. This also illustrates the importance of having the bubbles as close as possible to the target region to maximize the heat rise only in the target region.

Figure 9(a) and (b) show the instantaneous pressure contour for HIFU both in the absence and presence of microbubbles. Comparison of the two figures clearly shows the effect of the microbubbles on the wave, with the bubbles absorbing and scattering the incident acoustic wave. The lack of penetration of the waves in the bubble cloud region, i.e. beyond Z=40 mm, is evident. In addition to scattering, the bubbles are seen to emit pressure waves due to their own oscillations and this results in higher harmonics being produced. The production of these higher harmonics can be observed in Figure 10(a), which shows the variations of the pressure with time at Z=37 mm on the axis of the transducer. In the absence of bubbles, the signal is just sinusoidal at the fundamental HIFU frequency of 2.2 MHz. When bubbles are present, higher frequencies (4.4 MHz and 6.6 MHz) are also present and are superposed onto the 2.2 MHz. This is further illustrated in the frequency domain as shown in Figure 10(b). Two additional frequencies with significant energy can be seen at the harmonics: 4.4 MHz and 6.6 MHz. The amplitude of the peak at the higher harmonics is about 0.6 GPa/Hz, which represents a significant fraction (~25%) of the energy at the fundamental frequency, 2.6 GPa/Hz. This energy present in the higher harmonics when $\alpha=1\times10^{-5}$ leads to additional heat deposition, since tissues absorb energy at high frequencies preferentially (Holt & Roy 2001).

*Contribution of each Heat Source Terms*

In order to ascertain the individual contributions of the three heat source terms (ultrasound, ultrasound plus bubble acoustics, bubble viscous damping), the temperature distribution accounting for each source term separately is calculated and presented in Figure 11. The curve 'ultrasound only' shows the temperature distribution along the transducer axis at the end of 60 s of insonation considering only the ultrasound source in the absence of bubbles. A maximum temperature rise of less than 5 K is obtained at the geometric focal point. The second curve 'ultrasound plus bubble acoustics' shows the temperature distribution due to the combined ultrasound source and bubbles' acoustic emission/scattering. It is difficult to separate the contribution of the bubble acoustics from the primary acoustic field since both are inherently coupled and are estimated from the strain rate. It can be seen that the bubble acoustics' contribution is negligible here compared to the primary acoustic excitation contribution. Actually, the bubbles attenuate the primary ultrasound and lead to a reduction in the temperature rise and a slight increase in the pre-focal heating. The maximum temperature rise along the axis in this case reduces to 3 K and the peak heating region shifts to Z=35 mm, which is well ahead of the geometric focal region (i.e. closer to the acoustic source). The main contribution of heat deposition, however, comes from the viscous damping of the bubbles as illustrated in the third curve in Figure 11. A maximum temperature rise of 50 K is obtained when the bubbles' viscous damping is considered. The influence of this term is primarily limited to the vicinity of the bubbles even though conduction can heat nearby regions but to a much lesser extent. The temperature distribution now becomes bi-modal due to the presence/absence of bubbles near the axis. This bi-modal distribution in the temperature profile has also been observed in the numerical study of Okita et al. (2013). This also indicates the importance of localizing the bubbles very close the focal region to minimize heat conduction away from the bubbles leading to significant heating in the pre-focal regions.

*Effect of initial void fraction*

The effects of the initial void fraction on the heat deposition are illustrated by plotting the temperature profile along the axis for different initial void fraction values in Figure 12(a). For an initial void fraction $\alpha=10^{-6}$, a maximum temperature rise of 20 K is obtained. It is evident that the presence of bubbles leads to higher temperatures. This is accompanied by a shift of the location of the peak temperature towards the pre-focal region. These effects increase as the void fraction is increased. For $\alpha=10^{-5}$, the maximum heat deposition increases further to attain 50 K. At the same time, the attenuation in the pre-focal region is also larger, which results in the shifting of the peak towards the acoustic source by about 5 mm. Figure 12(b) shows a focal scan (locus of maximum pressure) along the axis. The pressure peak in the absence of bubbles has a value of 4 MPa, while it is significantly reduced when bubbles are present. For $\alpha=10^{-6}$, the attenuation causes the pressure peak to reduce to about 3 MPa. An even more significant attenuation results when the void fraction is increased. For example, for $\alpha=10^{-5}$, the attenuation is so significant that the peak in the focal region is no longer the largest pressure peak and is replaced by a secondary peak, which occurs at Z=35 mm.

*Importance of Bubble Localization*

As shown in the previous sections, the presence of bubbles can help by concentrating significant heat deposition near the bubble cloud. Hence the location of the bubbles during insonation plays a vital role and can be selected to improve the sought results near the target. To investigate this, we first consider what happens when the volume of the cylindrical region where the bubbles are present is reduced. Figure 13 shows the effect of the reducing the bubble cloud size for $\alpha = 10^{-5}$. The results with the bubbles distributed in a cylindrical volume of length 5 mm and diameter 5 mm, are compared with the experimental configuration of length 10 mm and diameter 10 mm cylinder. In both these cases, the geometric focus lies at the center of the cylinder. i.e. the base of the 10 mm cylinder is at an axial location of 35 mm and the base of the 5mm cylinder is at an axial location of 37.5 mm from the acoustic source. With the 5 mm cylinder, the bubbles are present only near the focus, thereby the attenuation of the ultrasound by the bubbles before reaching focal region is reduced leading to a higher temperature increase. When the bubbly region occupies the 10 mm cylinder, the maximum temperature rise obtained is 50 K and the axial location of the maximum temperature is at 35.5 mm. When the bubbles are distributed inside a 5 mm cylinder instead, the maximum temperature rise jumps to approximately 200 K and the location of maximum temperature is shifted closer to the geometric focus, i.e to 38 mm. Thus, localizing the bubbles closer to the focus leads to a significant increase in heat deposition.

The effect of location of the bubble cloud with respect to the geometric focus is illustrated in Figure 14. The position of the 5 mm cylinder is here shifted such that the base of the cylinder is now at an axial location of 38.75 mm. The temperature distribution along the axis obtained with the cylinder base at 38.75 mm is compared with the results obtained with the cylinder base at 37.5 mm. When the bubble cloud is closer to the geometric focus, the heat depostiion is further increased, again due to the reduced attenuation of ultrasound before it reaches the focus. A maximum temperature of 500 K is now obtained at an axial location of 39 mm. i.e the peak shifts approximately by the distance by which the bubble cloud is shifted.

The above two parameters: size of the cloud and its location illustrate the potential for control of the heat deposition via the injection of a bubble cloud near the tumor to ablate.

## Conclusions

Microbubble enhanced HIFU is studied numerically using an Eulerian-Lagrangian model. The non-linear acoustic field modeled in an Eulerian framework is coupled to a bubble dynamics solver, where the individual bubbles are tracked in a Lagrangian framework. The method is first validated using in vitro experiments in the absence of microbubbles. Good agreement for the temperature profile is obtained in and around the focal region. The method is then applied to a HIFU simulation in the presence of microbubbles. A reasonable qualitative agreement of the temperature profiles around the focal region is obtained and would be improved with finer grid resolution. Quantitative validation of the method is hampered at this point by the lack of exact details from the experiment. The individual contributions due to bubble acoustics and viscous damping are quantified. For the considered configuration, the effect of viscous damping of bubble oscillations is the primary contributor to the enhanced heat deposition when bubbles are present. The effect of initial void fraction on the temperature distribution is then illustrated and enhanced attenuation of the primary ultrasound and pre-focal heating is demonstrated at higher void fractions. The effect of the size of the bubble cloud and of its localization is then demonstrated by concentrating the bubbles in a smaller volume and closer to the focal location, which results in higher temperature increase due to reduced attenuation of the ultrasound.

## Acknowledgments

We gratefully acknowledge the support from NIH under SBIR Grant No. 1R43CA213866-01A1.

# FIGURE CAPTION LIST

Figure 1: Illustration of bubble volume spreading for void fraction computation using a Gaussian Kernel of characteristic radial extent, $\lambda$.

Figure 2: Grid sensitivity study. (a) Variation of the pressure history at the focal point for different number of discretization points per wavelength. 1.1 MHz transducer and a source pressure of 0.01 MPa. (b) Error in the solution for the peak pressure and its timing versus the number of discretization points per wavelength.

Figure 3: Characterization of the acoustic field in linear and non-linear regime in water. (a) Comparison of focal scan data along the axis with the experiments in Huang (2002) for a 1 MHz transducer with a low pressure amplitude, $P_a = 0.01$ MPa, corresponding to the linear regime. (b) Comparison of time history of pressure at the focus with the experiments in Canney et al. (2008) for a 2.2 MHz transducer with pressure amplitude $P_a = 0.28$ MPa, corresponding to the non-linear regime. Profile shows non-linear steepening of acoustic waves.

Figure 4: (a) Instantaneous contours of HIFU pressures showing focusing of the wave at $Z = 60$ mm. (b) Contours of the time-averaged (over 0.01ms) heat source *($q_{US}$)* computed from the hydrodynamic field. Logarithm of heat release is plotted for clarity. The simulations are based on the experiments in Huang et al. (2004) with a 1 MHz transducer.

Figure 5: Comparison of the temperature history obtained from HIFU simulations in phantom tissue without bubbles with the experiments of Huang et al. (2004) for 1s of insonation followed by cooling for 4s. (a) Temperature at the focal point, $Z= 60$mm. (b) At a point away from the focus, at $X = 0.1$ mm and $Z = 60$ mm.

Figure 6: Schematic of the experimental HIFU setup in the experiments of Kajiyama et al. (2010) with microbubbles showing the dimensions of the transducer and the focal length. The shaded cylindrical region of dimensions 10 mm x 10 mm contains a random distribution of microbubbles of diameter 1.3 $\mu$m.

Figure 7: Comparison of the temperature rise obtained in the focal region of simulations with the experiments of Kajiyama et al. (2010) in absence of bubbles (red lines) and with bubbles, $\alpha=1\times10^{-5}$ (black lines). The dashed lines represent computed temperature evolution at several axial locations indicated on the curve. The solid line represents spatial averages of the computes temperatures in the range 40mm<Z<45mm. Symbols are from the experiments (Kajiyama et al. 2010).

Figure 8: Contours of temperature rise at the end of 60s of insonation: (a) without bubbles, and (b) with bubbles ($\alpha=1\times10^{-5}$). The maximum temperature region in the absence of bubbles is along the axis of the transducer (X=0), but is altered drastically in the presence of bubbles.

Figure 9: Instantaneous pressure contours showing pressure wave focusing (a) without bubbles and (b) with bubbles ($\alpha=1 \times 10^{-5}$). The bubble sizes have been artificially enlarged by 200 times for clarity.

Figure 10: (a) Time history of pressure on at z=37mm along the axis, showing the presence of higher harmonics when $\alpha=1 \times 10^{-5}$ (b) FFT of pressure history at z=37mm along the axis. The peaks at the higher harmonics when $\alpha=1 \times 10^{-5}$ are approximately 0.6 GPa/Hz which represents a significant fraction of the energy present at the fundamental frequency, 2.6 GPa/Hz.

Figure 11: Contributions of various heat sources to axial temperature distribution. The simulations with bubbles are carried out for a bubble void fraction, $\alpha=1 \times 10^{-5}$. The viscous damping of bubble oscillations leads to the most significant heat release among all the heat source contributions.

Figure 12: Effect of void fraction on temperature and pressure distribution along axis. (a) Temperature profile along the axis after 60 s of insonation. Higher void fraction leads to higher heat rise albeit in the pre-focal region. (b) Instantaneous focal scan along the axis at 0.1 ms. Presence of bubbles leads to significant attenuation and shifting of pressure peak to pre-focal region.

Figure 13: Effect of the size of the bubble cloud on the temperature profile along the axis. The microbubbles are distributed inside a cylindrical volume of different dimensions. of the smaller cylindrical volume with the same void fraction brings the heating closer to the geometric focus and leads to higher heat deposition due to reduced attenuation of the ultrasound in the pre-focal region.

Figure 14: Effect of the localization of the bubble cloud on the temperature profile along the axis. For the same void fraction and same volume of the cylindrical bubbly region moving the cylinder closer to the geometric focus leads to higher heat deposition due to reduced attenuation of the ultrasound in the pre-focal region.

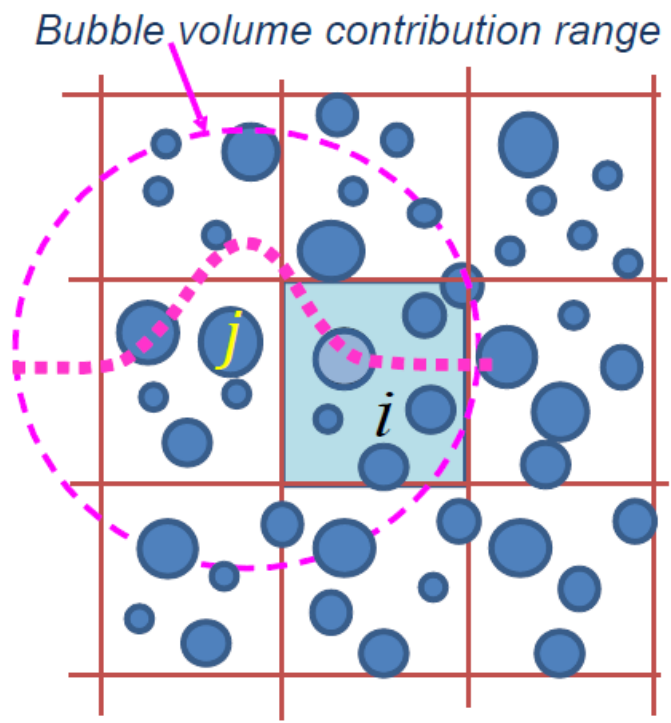

Figure 1: Illustration of bubble volume spreading for void fraction computation using a Gaussian Kernel of characteristic radial extent, λ.

(a)

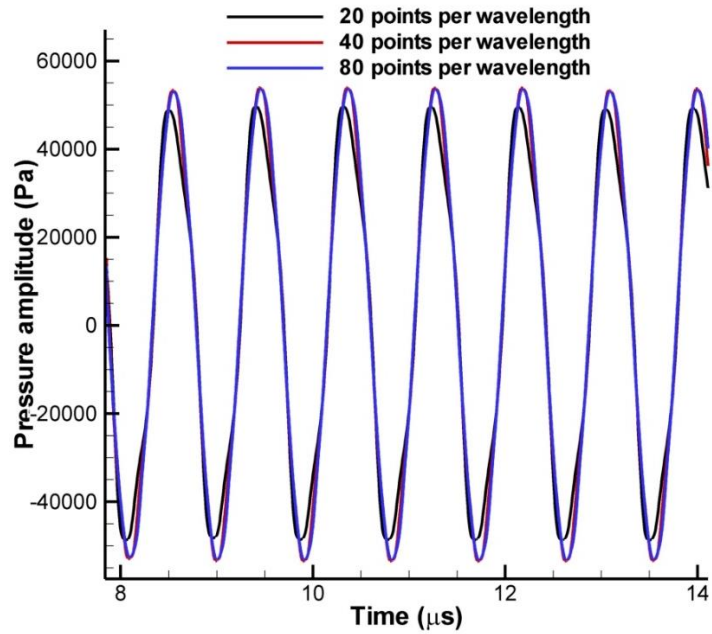

(b)

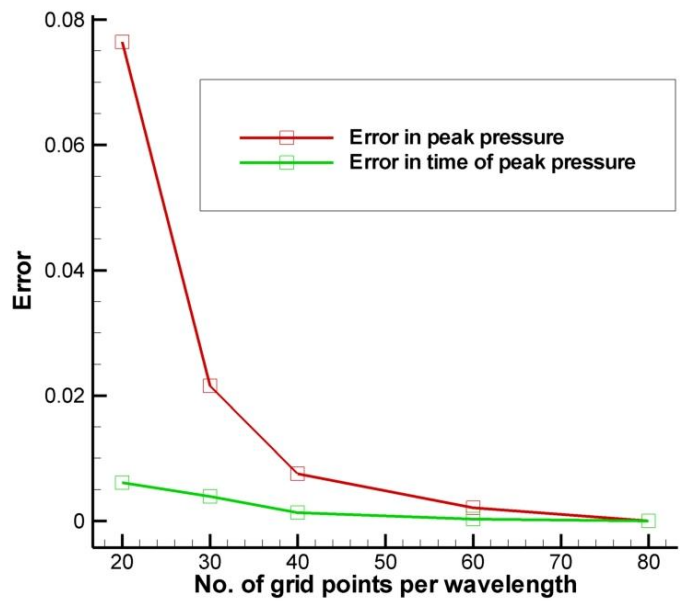

Figure 2: Grid sensitivity study. (a) Variation of the pressure history at the focal point for different number of discretization points per wavelength. 1.1 MHz transducer and a source pressure of 0.01 MPa. (b) Error in the solution for the peak pressure and its timing versus the number of discretization points per wavelength.

(a)

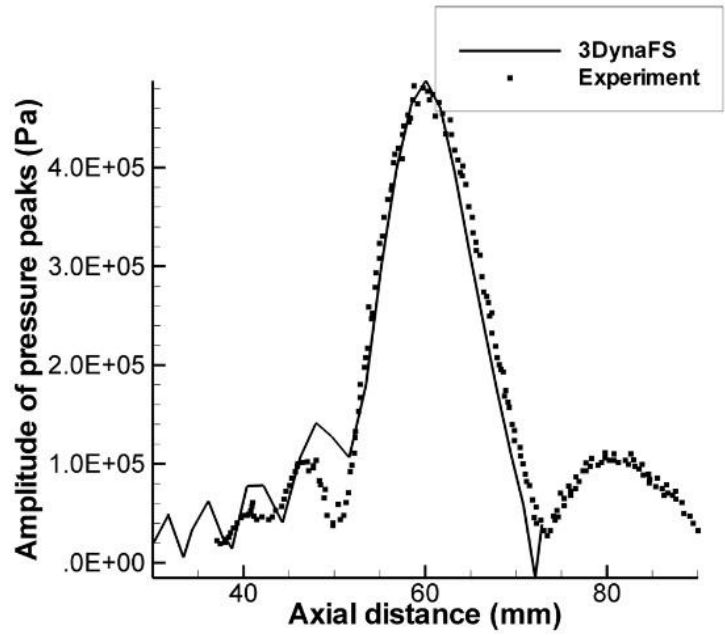

(b)

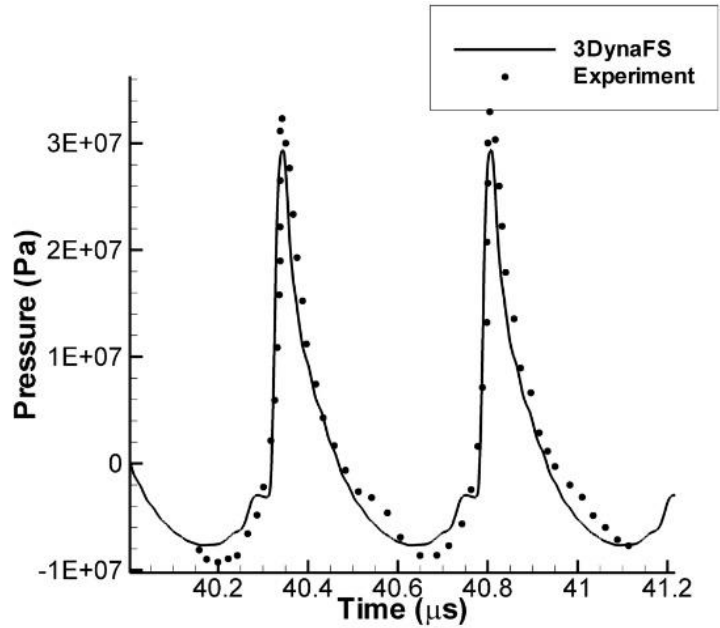

Figure 3: Characterization of the acoustic field in linear and non-linear regime in water. (a) Comparison of focal scan data along the axis with the experiments in Huang (2002) for a 1 MHz transducer with a low pressure amplitude, $P_a = 0.01$ MPa, corresponding to the linear regime. (b) Comparison of time history of pressure at the focus with the experiments in Canney et al. (2008) for a 2.2 MHz transducer with pressure amplitude $P_a = 0.28$ MPa, corresponding to the non-linear regime. Profile shows non-linear steepening of acoustic waves.

(a)

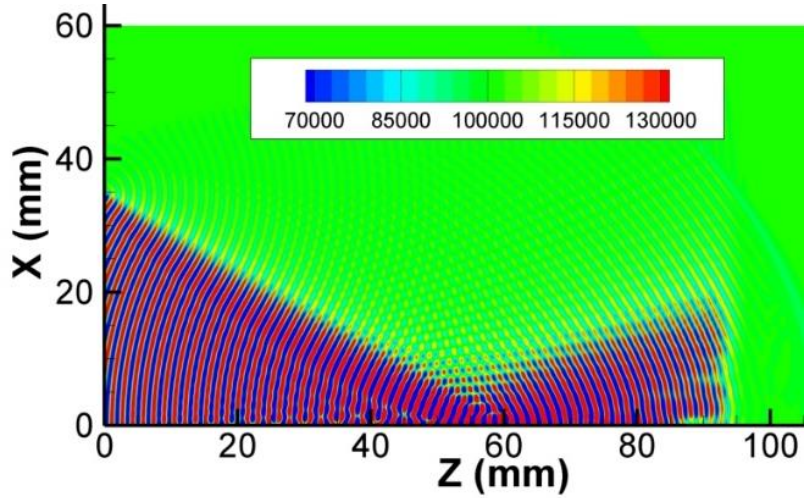

(b)

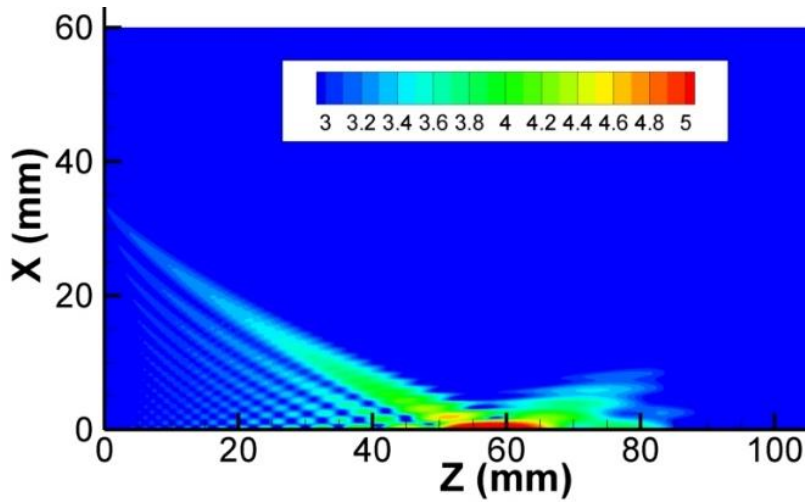

Figure 4: (a) Instantaneous contours of HIFU pressures showing focusing of the wave at $Z = 60$ mm. (b) Contours of the time-averaged (over 0.01ms) heat source $(q_{US})$ computed from the hydrodynamic field. Logarithm of heat release is plotted for clarity. The simulations are based on the experiments in Huang et al. (2004) with a 1 MHz transducer.

(a)

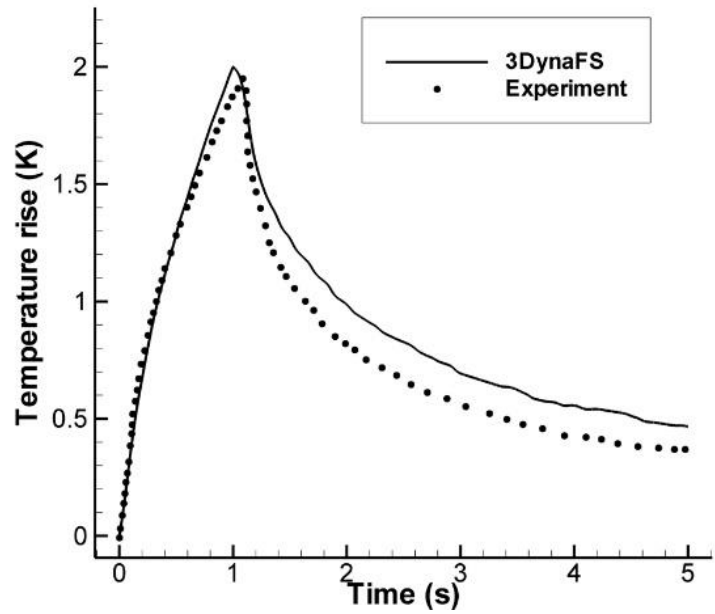

(b)

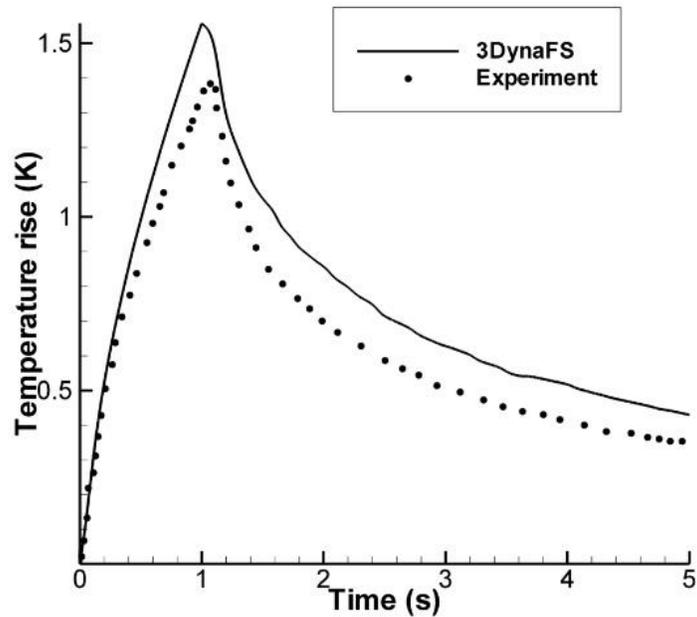

Figure 5: Comparison of the temperature history obtained from HIFU simulations in phantom tissue without bubbles with the experiments of Huang et al. (2004) for 1s of insonation followed by cooling for 4s. (a) Temperature at the focal point, Z= 60mm along the axis. (b) At a point away from the focus, at X = 0.1 mm and Z = 60 mm.

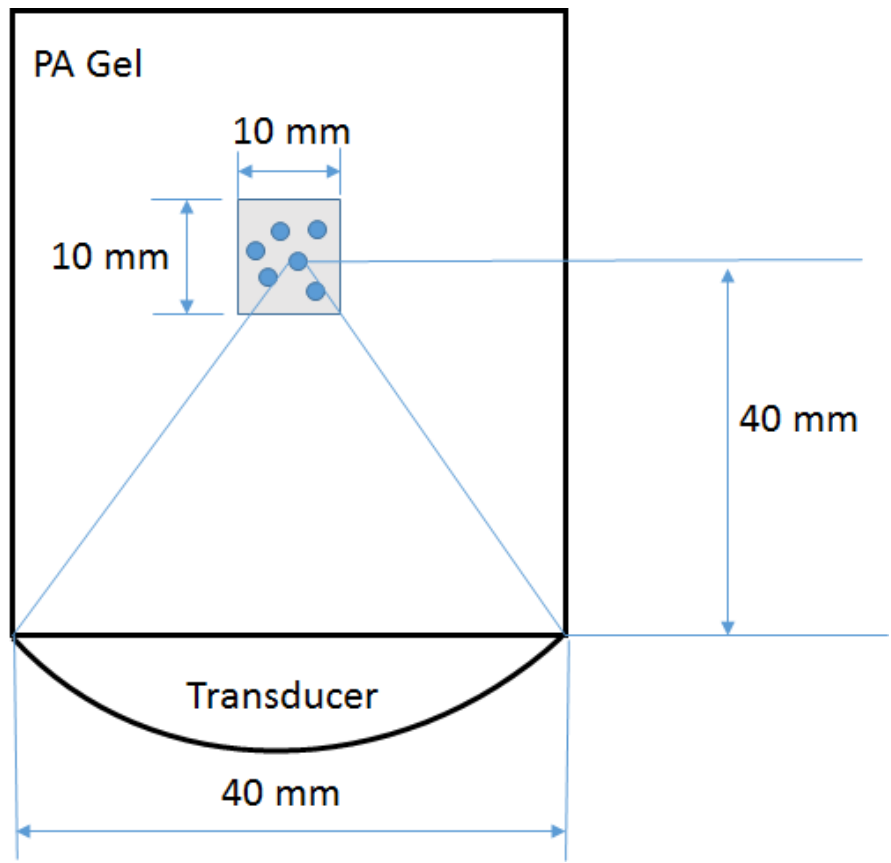

Figure 6: Schematic of the experimental HIFU setup in the experiments of Kajiyama et al. (2010) with microbubbles showing the dimensions of the transducer and the focal length. The shaded cylindrical region of dimensions 10 mm x 10 mm contains a random distribution of microbubbles of diameter 1.3 μm.

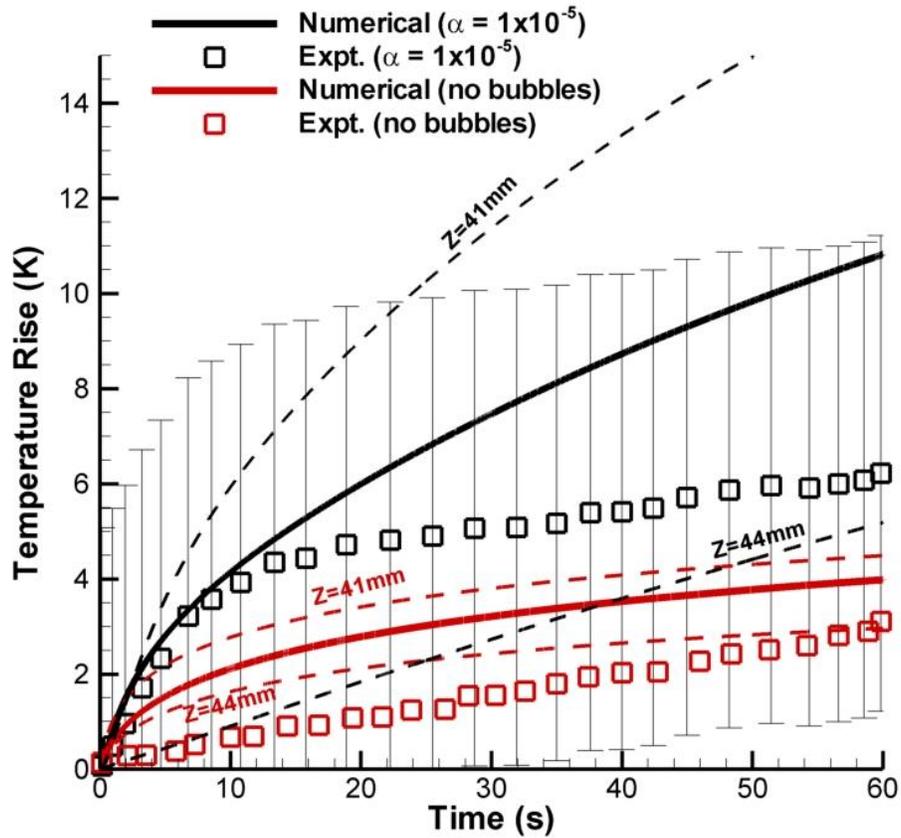

Figure 7: Comparison of the temperature rise obtained in the focal region of simulations with the experiments of Kajiyama et al. (2010) in absence of bubbles (red lines) and with bubbles, $\alpha=1\times10^{-5}$ (black lines). The dashed lines represent computed temperature evolution at several axial locations indicated on the curve. The solid line represents spatial averages of the computes temperatures in the range 40mm<Z<45mm. Symbols are from the experiments (Kajiyama et al. 2010).

(a)

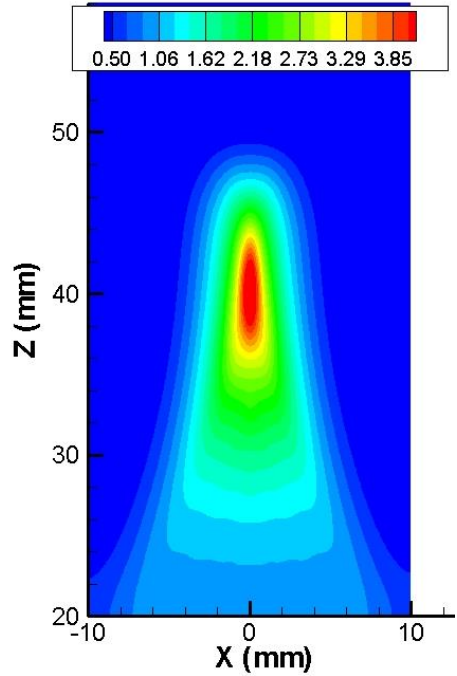

(b)

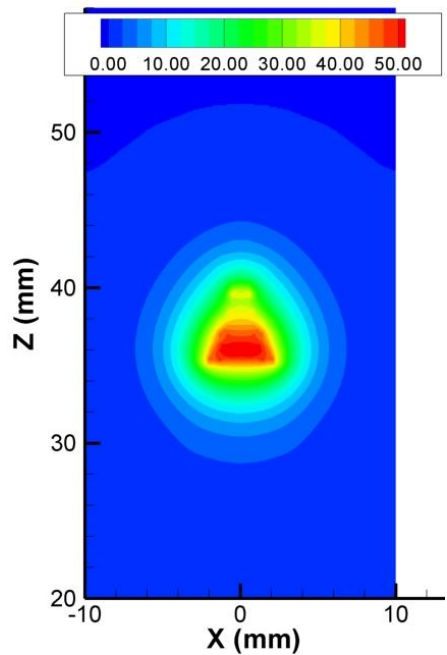

Figure 8: Contours of temperature rise at the end of 60s of insonation: (a) without bubbles, and (b) with bubbles ($\alpha=1\times10^{-5}$). The maximum temperature region in the absence of bubbles is along the axis of the transducer (X=0), but is altered drastically in the presence of bubbles.

(a)

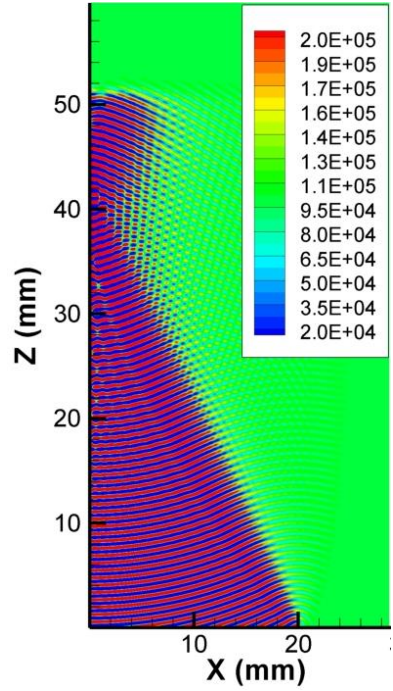

(b)

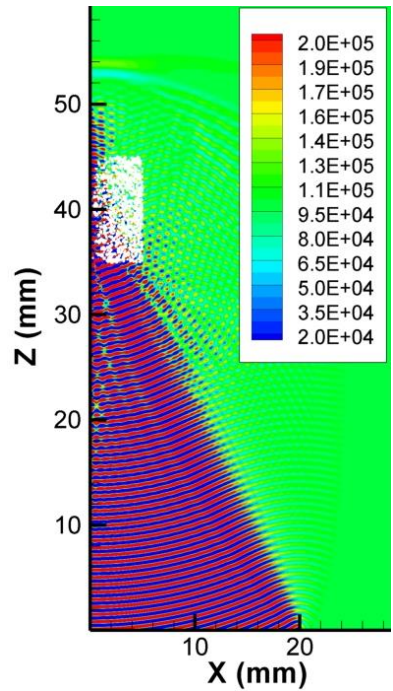

Figure 9: Instantaneous pressure contours showing pressure wave focusing (a) without bubbles and (b) with bubbles ($\alpha=1\times10^{-5}$). The bubble sizes have been artificially enlarged by 200 times for clarity.

(a)

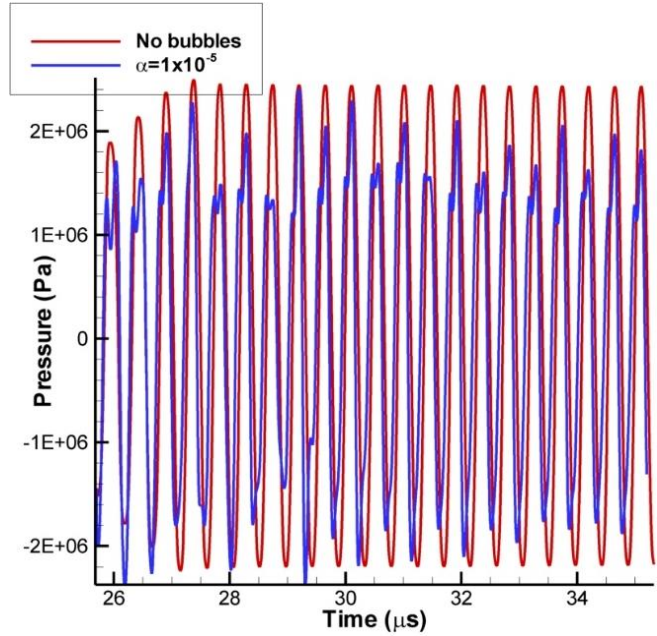

(b)

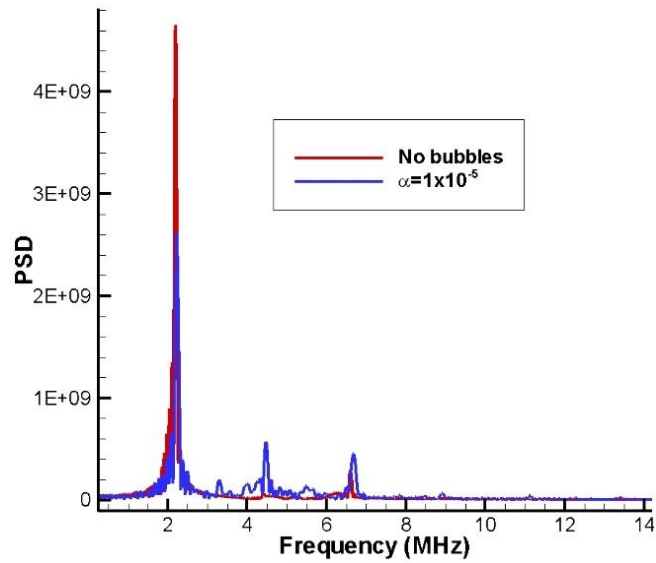

Figure 10: (a) Time history of pressure on at z=37mm along the axis, showing the presence of higher harmonics when $\alpha=1\times10^{-5}$ (b) FFT of pressure history at z=37mm along the axis. The peaks at the higher harmonics when $\alpha=1\times10^{-5}$ are approximately 0.6 GPa/Hz which represents a significant fraction of the energy present at the fundamental frequency, 2.6 GPa/Hz.

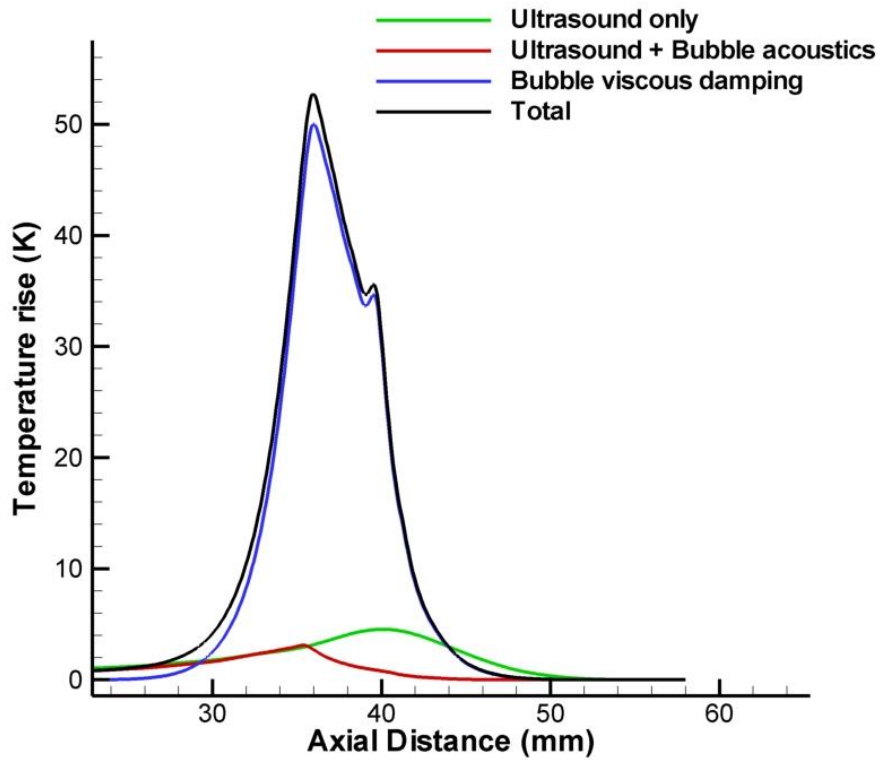

Figure 11: Contributions of various heat sources to axial temperature distribution. The simulations with bubbles are carried out for a bubble void fraction, $\alpha=1\times10^{-5}$. The viscous damping of bubble oscillations leads to the most significant heat release among all the heat source contributions.

(a)

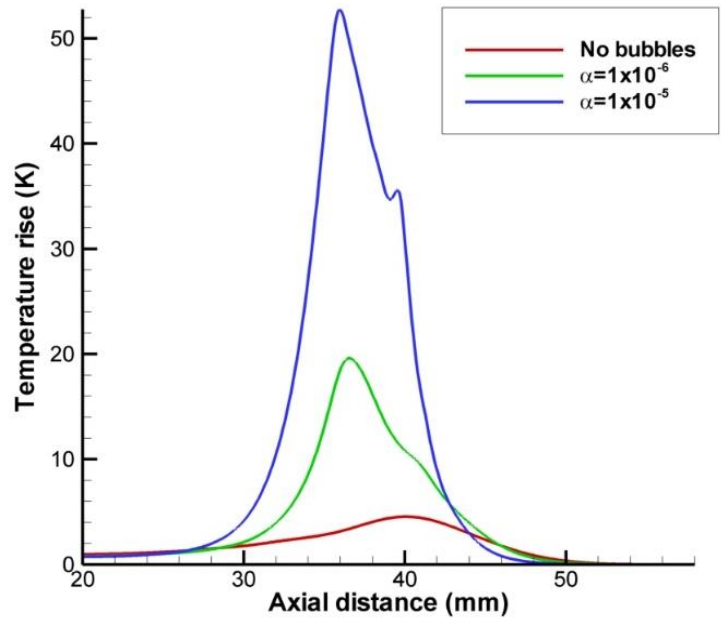

(b)

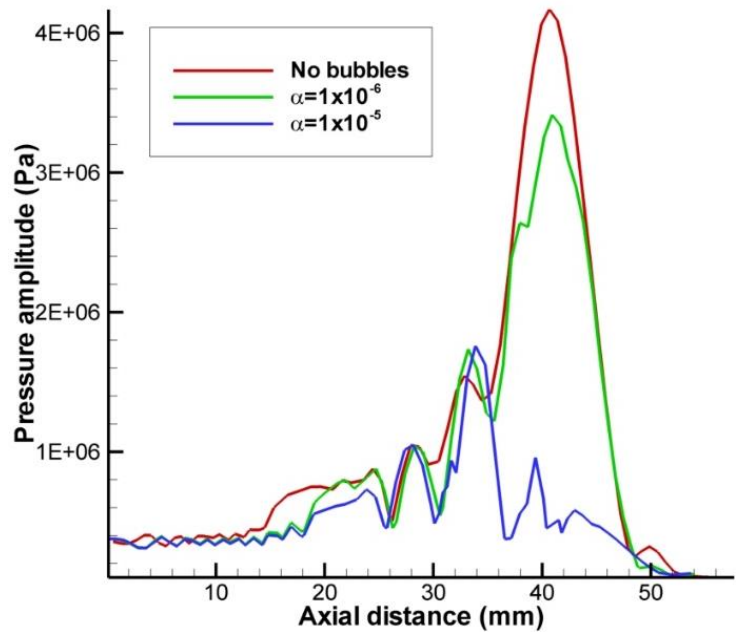

Figure 12: Effect of void fraction on temperature and pressure distribution along axis. (a) Temperature profile along the axis after 60 s of insonation. Higher void fraction leads to higher heat rise albeit in the pre-focal region. (b) Instantaneous focal scan along the axis at 0.1 ms. Presence of bubbles leads to significant attenuation and shifting of pressure peak to pre-focal region.

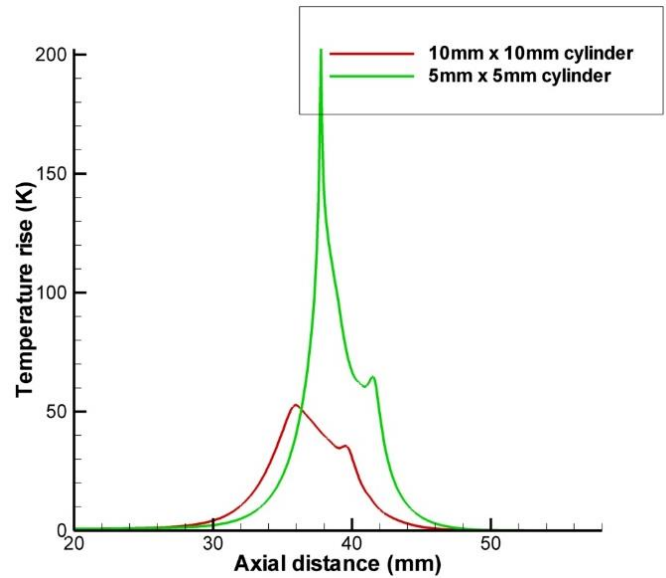

Figure 13: Effect of the size of the bubble cloud on the temperature profile along the axis. The microbubbles are distributed inside a cylindrical volume of different dimensions. of the smaller cylindrical volume with the same void fraction brings the heating closer to the geometric focus and leads to higher heat deposition due to reduced attenuation of the ultrasound in the pre-focal region

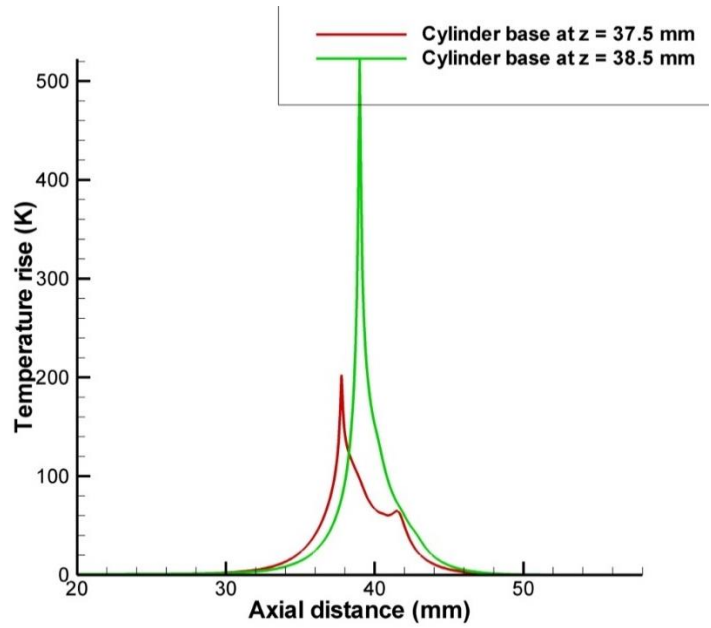

Figure 14: Effect of the localization of the bubble cloud on the temperature profile along the axis. For the same void fraction and same volume of the cylindrical bubbly region moving the cylinder closer to the geometric focus leads to higher heat deposition due to reduced attenuation of the ultrasound in the pre-focal region.